\documentclass[a4paper]{article}

\usepackage{atmohead2013}
\usepackage[english]{babel}
\usepackage[switch]{lineno}
\title{Data quality monitoring in the presence of aerosols and other
   adverse atmospheric conditions with H.E.S.S.}

\shorttitle{AtmoHEAD 2013: H.E.S.S. Atmosphere Data Quality}

\authors{
J. Hahn$^{1}$,
R. de los Reyes$^{1}$,
K. Bernl\"ohr$^{1}$,
P. Kr\"uger$^{1,2}$,
Y.T.E. Lo$^{3}$,
P.M. Chadwick$^{3}$,
M.K. Daniel$^{3,4}$,
C. Deil$^{1}$,
H. Gast$^{1,5}$,
K. Kosack$^{6}$,
V. Marandon$^{1}$
}

\afiliations{
$^1$ Max-Planck-Institut f\"ur Kernphysik, P.O. Box 103980, D 69029, Heidelberg, Germany \\
$^3$ University of Durham, Department of Physics, South Road, Durham DH1 3LE, U.K. \\
$^6$ CEA Saclay, F-91191 Gif-sur-Yvette, Cedex, France \\
$^2$ now at: Centre for Space Research, North-West University, Potchefstroom 2520, South Africa \\
$^4$ now at: University of Liverpool, Liverpool, L69 7ZE. U.K.\\
$^5$ now at: RWTH Aachen University, Physikzentrum, D 52056, Aachen, Germany  \\
}

\email{joachim.hahn@mpi-hd.mpg.de}

\abstract{Cherenkov telescope experiments, such as H.E.S.S., have been very
successful in astronomical observations in the very-high-energy (VHE; E $>$ 100 GeV) regime.
As an integral part of the detector, such experiments use Earth's
atmosphere as a calorimeter. For the calibration and energy determination,
a standard model atmosphere is assumed. Deviations of the real atmosphere
from the model may therefore lead to an energy misreconstruction of
primary gamma rays.
To guarantee satisfactory data quality with respect to difficult
atmospheric conditions, several atmospheric data quality criteria are
implemented in the H.E.S.S. software. These quantities are sensitive to
clouds and aerosols.
Here, the \emph{Cherenkov transparency coefficient} will be presented. It is a
new monitoring quantity that is able to measure long-term changes in the
atmospheric transparency. The Cherenkov transparency coefficient derives
exclusively from Cherenkov data and is quite hardware-independent.
Furthermore, its positive correlation with independent satellite
measurements, performed by the Multi-angle Imaging SpectroRadiometer
(MISR), will be presented.}


\begin{document}
\maketitle

%Begin a section.
\section{Introduction}
During the last three decades, imaging atmospheric
Cherenkov telescopes (IACTs) have qualified as powerful
instruments for gamma-ray astronomy in the very-high-energy (VHE; E $>$ 0.1 TeV) 
regime, allowing detailed studies of the most violent phenomena known in the Universe.
The gamma-ray flux at these energies is rather low and the
IACT technique provides the large effective areas required
making use of telescopes on the ground. Due to its opacity
 at these energies, the Earth's atmosphere acts as the
calorimeter of the detector system; therefore, the VHE
photons can be observed only indirectly at ground level.
One of the main strengths of this type of detector is its low
energy threshold, which unfortunately increases with the
atmospheric absorption.
The atmospheric absorption will also affect the reconstruction
 of the energy of the primary particle, since shower
images are compared to Monte Carlo shower simulations
 for which nominal hardware parameters and average atmospheric
 conditions at the H.E.S.S. site (23$^\circ$16'18'' S, 16$^\circ$30'00'' E, 1800 m a.s.l)
 are assumed \cite{bib:Aharonian2006}\cite{bib:Bernlohr2000}. However, this comparison might be affected by changes in the
telescope efficiency, which include not only changes in
the optical efficiency and photo-sensor response but also
atmospheric fluctuations. Any atmospheric phenomenon
that acts as a light absorber will attenuate
Cherenkov light from EAS (Extensive Air Shower) particles
 and therefore reduce the amount of Cherenkov photons
 that reach the detector, which will cause an underestimation
 of the energy of the primary gamma ray. This is especially
problematic for spectral analysis, since mis-reconstructed
energies lead to biased values of the flux normalization
and, in particular, in the case of non-power-law spectra,
other spectral parameters \cite{bib:Nolan}.
To limit such effects to a minimum, corresponding monitoring quantities have to be used in the Cherenkov technique in order to detect data that are taken in the presence of clouds and aerosols\footnote{The technical framework presented in this document is specific to the H.E.S.S. Analysis Program in Heidelberg (hap-HD).}. 
These data quality quantities are applied in an automated and uniform way to
the data set. This guarantees a well-defined and reproducible data selection which is especially
important for the analysis of large amounts of data, as for example in surveys.
We will introduce the most important atmospheric conditions that affect
spectral shower reconstruction.
Furthermore, we want to present a new way to estimate
 the atmospheric transparency by using only observables
 and calibration parameters from the Cherenkov data
 taken with the H.E.S.S. telescope array and a detailed comparison
of this new atmospheric monitoring quantity with MISR
(Multi-angle Imaging SpectroRadiometer) satellite data.
Finally, the last part will contain a short systematic study
on the effect of the atmospheric transparency on reconstructed
 spectral parameters.

\section{Atmospheric Effects And Their Detection}

\subsection{Clouds}
\label{clouds}
The maximum of the Cherenkov emission from air showers triggered by particles of energies within the H.E.S.S. energy domain (E $\ge$ 300 GeV) is at altitudes between $\sim$6-11 km (see~\cite{bib:Bernlohr2000}). 
Therefore, any atmospheric light-absorbing structure situated at or below such altitudes might
at least partially attenuate the Cherenkov light from the shower. As a result, fewer 
photons reach the cameras, which leads to a decreased trigger probability and ultimately
lower single telescope and central trigger rates \cite{bib:Funk2004}.

If absorbing structures (local clouds) are passing through the field-of-view, a fluctuating behavior in the central trigger rate\footnote{Here, all trigger rates are assumed to be corrected for the decrease that goes with increasing the observational zenith-angle.}
on time-scales smaller than the standard duration of data sets (typically, a 28-min \emph{run}) can be observed. 
Observations taken under such atmospheric conditions
can be identified by
\begin{itemize}

  \item (i) fluctuations in the central trigger rate, if caused by rather small-scale
  absorbers (e.g. small clouds)

  \item (ii) steady decline in the central trigger rate, if connected to large-scale 
  absorbing structures moving into the field-of-view (e.g. large clouds)

\end{itemize}

In order to quantify these changes in the central trigger rate for each data set taken with with H.E.S.S., the average central trigger rate is calculated over 10-s time intervals. The resulting evolution of the trigger rate is then fitted by a linear function. 

The slope of this function can then be used to address case (ii). As a data quality criterion, the extrapolated trigger rate, obtained by multiplying 
the slope of the trigger rate evolution with the run duration, is required to be within 30\% of the run-averaged value of 
the trigger rate.

Fluctuations of the central trigger rate (case (i)) can be quantified by the rms of the residuals of the fit. As an atmospheric data monitoring quantity in H.E.S.S., the rms value is divided by the time-averaged central trigger rate. The corresponding quality cut is at 10\%. 

\begin{figure}[h!!!]
  \begin{center}
  \includegraphics[width=0.49\textwidth]{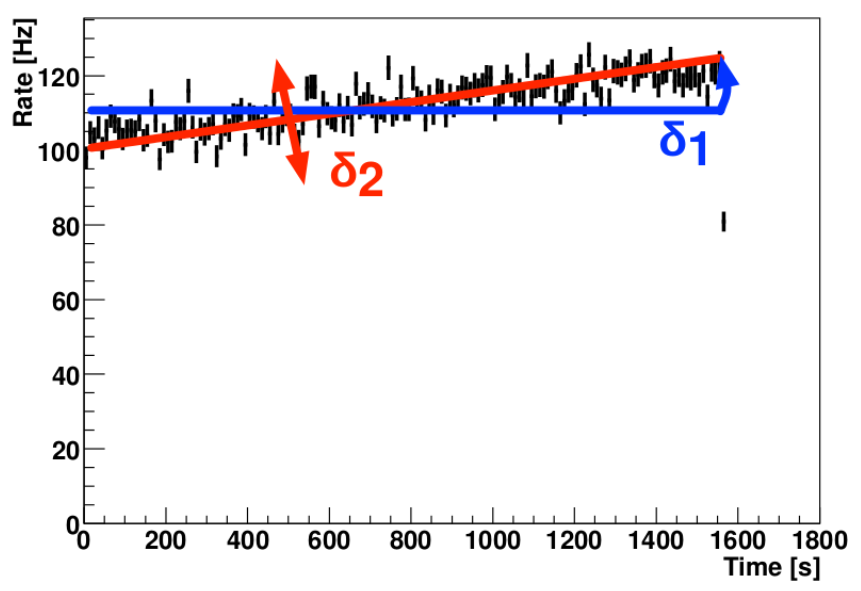}
  \caption{Behavior of the central trigger rate in the presence of clouds moving through the field-of-view. Fluctuations can be quantified by the rms of the data points with respect to a linear fit (here called $\delta_2$), a steady decline by the slope of the fit function (here called $\delta_1$).}
  \label{deltas}
  \end{center}
\end{figure}

The derivation of these quantities is also illustrated in Fig.\ref{deltas}.

However, these quantities are only sensitive to clouds that affect the central trigger rate on time-scales smaller than the run duration.

\subsection{The Cherenkov Transparency Coefficient}
The detection of modifications to the central trigger rates on time-scales much larger
than the duration of the individual data sets, as resulting from shower attenuation
by large-scale absorbing structures like aerosol layers, requires a different approach.

We have developed a new 
quantity, the \emph{Cherenkov transparency coefficient}, which is designed to be
as hardware-independent as possible in order to separate
hardware-related effects from the decrease in trigger rates
caused by large-scale atmospheric absorption.

For the definition of the Cherenkov transparency coefficient we assume that the zenith-angle-corrected single telescope trigger rates $R$ are dominated by cosmic-ray (CR) protons.
The local CR proton spectrum in the relevant energy range is approximately $f(E) = 0.096\cdot (E/\textrm{TeV})^{-2.70} \textrm{m}^{-2} \textrm{s}^{-1}\textrm{TeV}^{-1}\textrm{sr}^{-1}$ \cite{bib:BESS98}.
Hence, the trigger rates can be estimated by
\begin{eqnarray}
	R &\sim& \int_{0}^\infty A_{\rm eff}(E)f(E)\mathrm{d}E\\ 
	  &\simeq& k\cdot E_0^{-1.7 + \Delta},
\end{eqnarray}
 
where $E_0$ is the energy threshold  and $A_{\rm eff}$ the effective area of the telescopes. The term $\Delta$ allows one
to take into account higher-order corrections, such as energy-dependent shower profiles. 
Furthermore, $E_0$ is assumed to be inversely proportional to the average
 pixel gain $g$ \cite{bib:Aharonian2004}, the muon efficiency $\mu$ \cite{bib:HESSMuons} and the atmospheric
 transparency, parametrized by a factor $\eta$ so that
$ E_0 \propto (\eta \cdot \mu \cdot g )^{-1}$. The quantities $\mu$ and $g$ are
telescope-specific, so for each telescope $i$ one can derive an estimation of $\eta$,
$$\eta \propto \frac{R_i^{\frac{1}{1.7-\Delta}}}{\mu_i\cdot g_i} \equiv t_i.$$
Random fluctuations in the trigger of a single telescope
are removed by selecting only those events where at least two telescopes are triggered in coincidence.
The corresponding trigger rate will therefore depend on the number
of active telescopes, so the averaged trigger over all $N$ active telescopes
 is calculated and rescaled by a factor of $k_N$
that depends on the telescope multiplicity. The corresponding
values are $k_3 = 3.11$ for observations with three participating telescopes 
and $k_4 = 3.41$ for observations with four telescopes. This rescaling also cancels
out the contribution of other CR species to the trigger rate.

The Cherenkov transparency coefficient ($T$) is then defined as
$$T \equiv \frac{1}{N\cdot k_N}\sum_i{t_i}.$$

\begin{figure*}[!t]
  \centering
  \includegraphics[width=\textwidth]{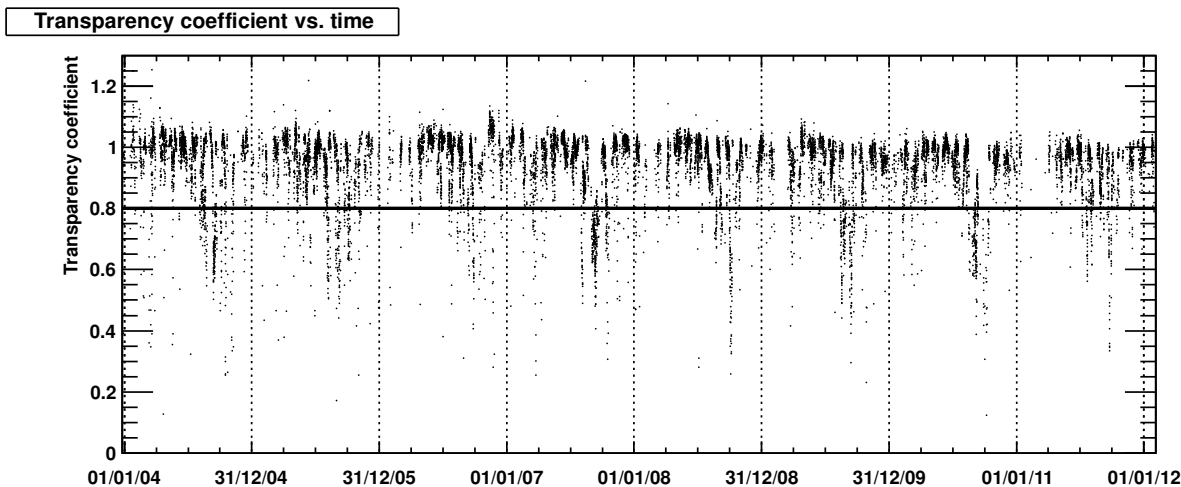}
  \caption{Evolution of the transparency coefficient over 
eight years of H.E.S.S. observations. The solid line indicates the current data quality cut
value at 0.8 \cite{bib:Hahn2014}. The distribution is sharply peaked at 1 with
a FWHM $\sim 9\%$.}
  \label{TransEvo}
 \end{figure*}

Figure \ref{TransEvo} shows the evolution of the Cherenkov transparency coefficient
over a time of eight years of H.E.S.S. observations. The quantity is strongly 
peaked at unity (with a relative FWHM of $\sim 9\%$) and shows periodic downward-fluctuations
around September.

This quantity is used as the third and the last atmospheric data quality parameter with a cut value of 0.8 (see section \ref{Syseff}).
The cut affects $\sim$11\% of all runs. The cuts on trigger rate fluctuation and slope, as described in section \ref{clouds}, affect $\sim$6\% and $\sim$4\% of all runs, respectively.
It should be noted that there is a large overlap between the three cuts.

Runs affected by at least one of the three cuts are flagged as not fulfilling the quality standards for spectral reconstruction. However, they can still be used in the creation of images or source detection since for such purposes no spectral information is necessary.

\section{Atmospheric Absorption Due To Aerosols}
The sensitivity of the Cherenkov transparency coefficient
to the concentration of aerosols in the atmosphere will be
confirmed through a positive correlation with independent
aerosol measurements, such as satellite data. The influence
of the atmospheric aerosols on transparency is quite complex
 and strongly depends on the detailed scattering and absorption
 properties of the different aerosol types (e.g. sulphate,
 dust, organic carbon, sea salt) and their relative concentrations in the atmosphere at a given time. Many studies of the
atmospheric absorption of aerosols have been carried out,
not only for astronomical purposes but also for climate and
atmospheric studies. However, we are particularly interested
 in those related with increases of aerosol absorption due
to biomass burning. That is, agricultural biomass burning takes place
every year around September in Namibia and its neighboring countries. Aerosols from such processes seem
to decrease the amount of UV solar radiation reaching the
surface by up to 50\%, with typical values in the range of $\sim$
15-35\% \cite{bib:Kalashnikova2007}.

In the following, we will test for a correlation between
the Cherenkov transparency coefficient and the Aerosol
Optical Depth (AOD), or more specifically with the atmospheric
 transparency ($\propto \exp(-AOD)$).
The MISR (Multi-angle Imaging SpectroRadiometer)
instrument on board NASA's \emph{Terra} spacecraft has a better
spatial resolution (1.1 km in \emph{global mode}) \cite{bib:Diner1988} with
respect to other satellite instruments, and has the capability
to observe at different viewing angles, so that MISR can
distinguish between different types of atmospheric particles
 (aerosols), different types of clouds and different land
surfaces.
The processed (Level 3) AOD data used in this study
have proven to be in better agreement with the ground-based
 Aerosol Robotic Network (AERONET) measurements
 \cite{bib:Tesfaye2011} than previous satellite measurements. In particular, a detailed 10-year study of the aerosol climatology
 with MISR over South Africa, Namibia's neighbor
country, has revealed that the northern part of South Africa
seems to be rich in aerosol reservoirs and the aerosol concentration
 (based on optical depth) is 34\% higher than that
in the southern part of the country \cite{bib:Tesfaye2011}.

%\subsection{Correlation Between Cherenkov
%Transparency Coefficient And MISR Data}
Tesfaye et al. (2011) have also found seasonal changes in
the aerosol composition in South Africa. During summer
and early winter in the southern hemisphere, the northern
part of South Africa is dominated by a mixture of coarse-mode
 and accumulation-mode particles, which are a result
of air mass transport from arid/semi-arid regions of the 
central parts of South Africa, Botswana and Namibia. In the
time from August to October (winter and early summer) it
is dominated by sub-micron particles. The most important
sources of sub-micron particles are industrial and rural 
activities (including mining and biomass burning).
The periodic drops in the Cherenkov transparency 
coefficient for the H.E.S.S. site (see Fig. \ref{TransEvo}) correlate with the
seasonal increase of sub-micron particles due to, among
other sources, biomass burning like in nearby South
Africa. This gives an indication of the main atmospheric
phenomenon responsible for the reduced trigger rates of
some H.E.S.S. observations, especially in early summer,
and points to the Cherenkov transparency coefficient as
a good data quality parameter to monitor the atmosphere
transparency.
Therefore, we expect a strong and positive correlation
with the AOD measured by satellites. To do this, we used
the AOD retrieved from MISR data and the Cherenkov 
transparency coefficient from H.E.S.S. data. Both data sets 
cover the same period of time between 2004 and 2011. 
\begin{figure}[h!!!]
  \begin{center}
  \includegraphics[width=0.49\textwidth]{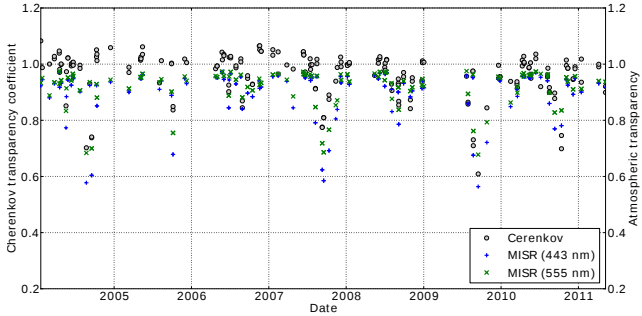}
  \caption{Cherenkov transparency coefficient measured in the
time interval 2004$-$2011, together with the
MISR atmospheric transparency measurements in 443 nm
(blue points) and 555 nm (green points).}
  \label{TimeLine}
  \end{center}
\end{figure}
\begin{figure}[h!!!]
  \begin{center}
  \includegraphics[width=0.49\textwidth]{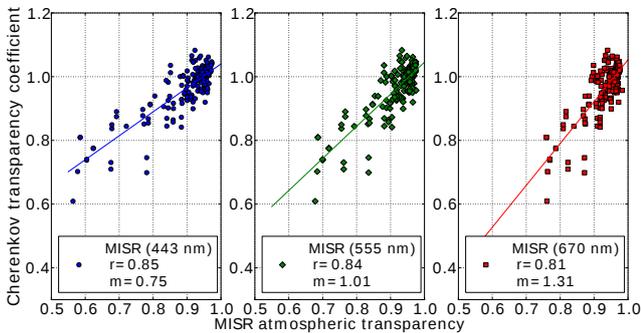}
  \caption{MISR atmospheric transparency ($\exp(-AOD)$) vs. the Cherenkov transparency coefficient. The three
wavelengths measured by the MISR satellite are represented 
in different colors: 443 nm, blue; 555 nm, green; and 670
nm, red. The resulting correlation is plotted as a solid line
with the corresponding color of the MISR wavelength.}
  \label{CorrPlot}
  \end{center}
\end{figure}

The
processed (level 3) MISR AOD data at the H.E.S.S. site
(with a grid spatial resolution of 0.5$^\circ$ x 0.5$^\circ$) at three 
different wavelengths (443 nm, 555 nm and 670 nm), from UV
to red wavelengths were used. Note that the satellite only
measures the AOD during daytime and, depending on latitude, 
the satellite samples a fixed location every 2 to 9 days.
The overlap of satellite measurements and H.E.S.S. data
taking is therefore sparse within a time interval of overlap
of 24 hours, reducing the available data set for the 
correlation study (only 2\% of the H.E.S.S. data can be used).
Fig. \ref{CorrPlot} quantifies the correlation between the 
atmospheric absorption ($\propto \exp(-AOD)$) for the three different 
wavelengths measured by MISR and the Cherenkov transparency 
coefficient. The solid lines are the results of a linear 
fit between measurements.
The Pearson correlation coefficients for the wavelengths
443 and 555 nm (blue and green) are $\sim$0.85 and $\sim$0.84 
respectively. This shows a positive and strong correlation 
between the atmospheric transparency measured from 
satellites and the Cherenkov transparency coefficient, in 
particular for the blue band, which is most relevant for this study
since the number of Cherenkov photons emitted per path
length in a certain wavelength range (eq. (1) in \cite{bib:Bernlohr2000}) is 
maximum in the UV-blue part of the spectrum.

Figure \ref{CorrPlot} also shows an increase of the steepness (“m” in
the figure) of the best fit of the linear correlation, with 
increasing wavelength. This is due to the fact that the 
atmospheric transparency measured with the MISR satellite 
decreases towards shorter wavelengths, while the Cherenkov
transparency coefficient is always the same.
The decrease of the atmosphere transmission with 
decreasing wavelength can be explained by simple Mie 
scattering. Tesfaye et al. (2011) established an inverse 
proportionality between the aerosol particle size and their 
extinction efficiency at a certain wavelength. An increase of
the AOD at short wavelengths therefore indicates the 
presence of sub-micron (radii $<$ 0.35 $\mu$m) particles, attributed
by the authors to urban pollution (sulphates) and extensive
biomass burning activities (carbonaceous aerosols).
As a consequence, the aerosol-induced reduction in
the atmosphere transparency is expected to be more 
pronounced at shorter wavelengths, which is where the bulk
of the Cherenkov light is emitted.

\section{Systematic Effect On Reconstructed Spectra}
\label{Syseff}
In order to investigate
systematic effects of the Cherenkov light attenuation of EASs by aerosols on the reconstructed gamma-ray spectrum, as traced by the Cherenkov transparency coefficient $T$, we have analyzed data taken on the Crab Nebula, a standard candle at
TeV energies without any detectable variability over timescales of years.

The full data set investigated was recorded during the years from 2004 to 2011 and has an exposure of 84 hours. In order to keep systematic effects arising from the radial acceptance profile in the cameras (see, e.g. \cite{bib:Aharonian2006}) to a minimum, only
observations within a one-degree offset from the source were used. Also, to minimize a possible zenith-angle-dependent
 energy bias, only data taken at zenith angles smaller than 47 degrees have been
selected. The data has been divided into subsets
 corresponding to different ranges of the atmospheric transparency parameter
 after applying standard quality
criteria to remove those runs with technical problems or
with small clouds in the field-of-view during the observations
 \cite{bib:Aharonian2006}. The standard cut-based analysis using simple
air shower image parameters, the Hillas analysis
\cite{bib:Hillas1985}, was then employed to obtain spectral information for
each subset.

The gamma-ray spectrum of the Crab Nebula has been measured by H.E.S.S. \cite{bib:Aharonian2006} and was found to have an approximate power-law shape with some curvature at the highest energies. For a pure power-law fit in the energy range ($0.41$-$40$) TeV, the flux normalization at $1 \textrm{TeV}$, $\phi_{0,\mathrm{Crab}}$, and the spectral index, $\Gamma_{\mathrm{Crab}}$, were found to be $\phi_{0,\mathrm{Crab}}=(3.45\pm0.05_{\mathrm{stat}}\pm0.69_{\mathrm{sys}})\times 10^{-11}\textrm{cm}^{-2}\textrm{s}^{-1}\textrm{TeV}^{-1}$ and $\Gamma_{\mathrm{Crab}}=2.63\pm0.01_{\mathrm{stat}}\pm0.10_{\mathrm{sys}}$.

Assuming that atmospheric absorption leads to an underestimation of the reconstructed energy by a constant attenuation factor, a power-law spectrum is expected to stay form invariant with changing atmospheric conditions. However, this shift in the spectrum energy range is expected to bias the estimated flux normalization at a given reconstructed energy. Quantitatively, assuming the reconstructed gamma-ray energy $E_{\mathrm{reco}}$ and the true energy $E_{\mathrm{\mathrm{true}}}$ to be related via $E_{\mathrm{reco}}\propto T\times E_{\mathrm{true}}$, 
one finds
\begin{eqnarray}
 \label{simpmod}
 \frac{\mathrm{d}F}{\mathrm{d}E_{\mathrm{true}}} \propto E_{\mathrm{true}}^{-\Gamma} ~~~~~\Leftrightarrow ~~~~~ \frac{\mathrm{d}F}{\mathrm{d}E_{\mathrm{reco}}} \propto E_{\mathrm{reco}}^{-\Gamma}\cdot T^{\Gamma-1}
\label{eqfit}
\end{eqnarray}

\begin{figure}[h!!!]
  \begin{center}
  \includegraphics[width=0.39\textwidth]{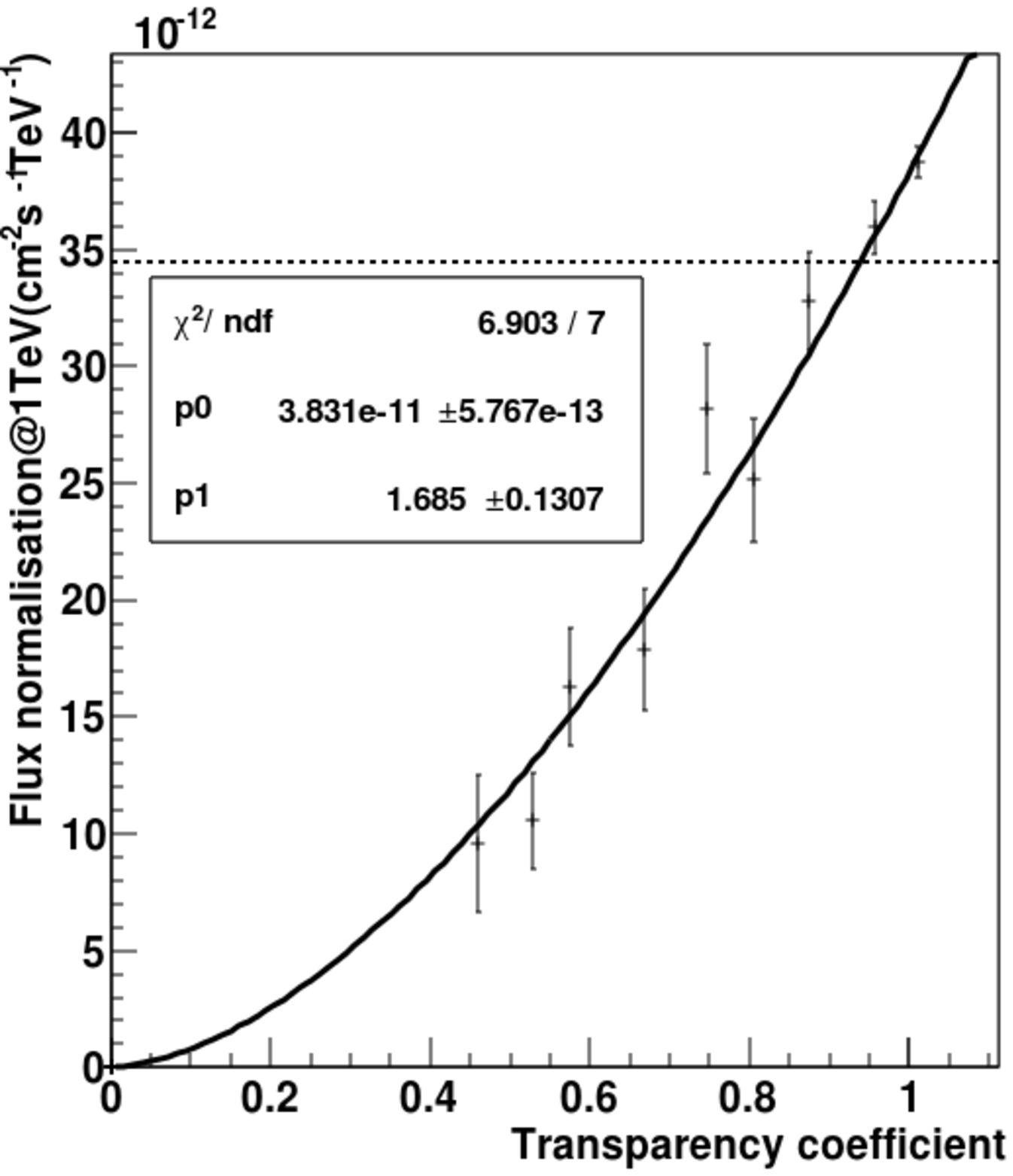}
  \includegraphics[width=0.39\textwidth]{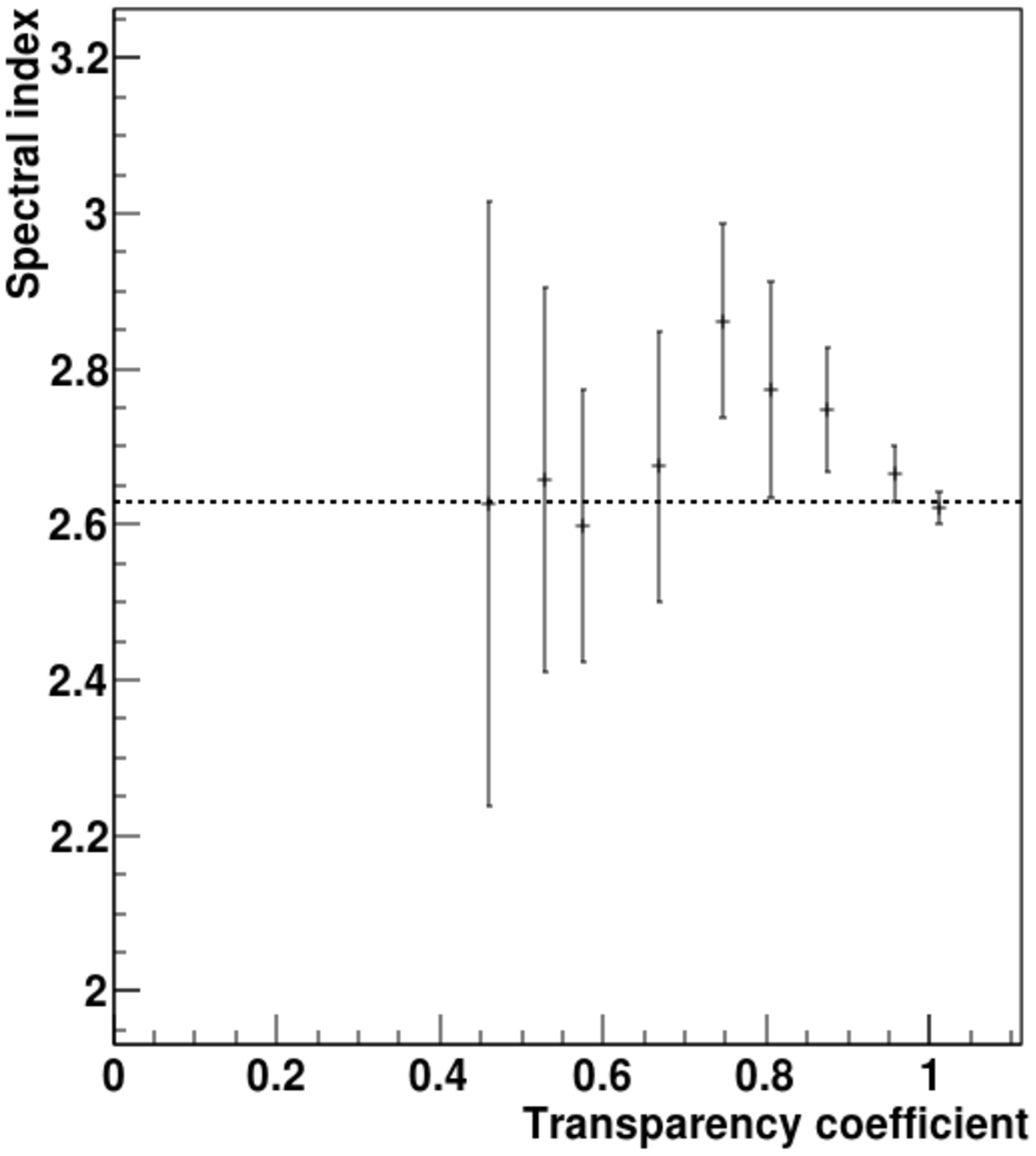}
  \caption{Flux normalization at 1 TeV (top) and spectral index (bottom) for Crab Nebula data taken during 8 years of H.E.S.S. operation. Abscissa values are given by the mean value of the transparency coefficient in the respective subset. In the top panel, best-fit values for a power-law model are shown for the flux normalization at T=1 (p0) and the exponent (p1). Dashed lines represent the published results \cite{bib:Aharonian2006}.}
  \label{SysStudy}
  \end{center}
\end{figure}

Fig. \ref{SysStudy} shows the reconstructed flux normalization as a function of $T$.
The data is well fitted by a power law with an exponent $\Gamma - 1 = 1.69 \pm 0.13$ and
confirms the $T$-dependence as expected from the simple model given by Eq. \ref{simpmod}.
The reconstructed spectral index $\Gamma$ shows no significant dependence on $T$.

The application of the presented data quality cuts limits the bias in reconstructed flux normalization to values smaller than 20\%.

\section{Conclusions}
H.E.S.S. uses an uniform and automatic data quality selection scheme featuring
 three atmospheric data quality quantities. 
These quantities are designed to identify data taken in the presence of atmospheric absorbers like clouds and aerosols, the former of which are relatively easy to detect by parametrising the time evolution of the central trigger rate within each recorded data set.

On the contrary, large-scale aerosol and cloud layers persist longer in the atmosphere; as a 
result, their signature on the trigger rate is of a larger time-scale than the typical 
data set duration.

In order to detect such absorber structures, a new and mostly hardware-independent quantity has been developed, the Cherenkov transparency coefficient $T$. Together with the other atmospheric data quality monitoring quantities,
its application as a cut quantity limits the bias in reconstructed flux normalization
to values smaller than 20\%.

This quantity is sensitive to elevated aerosol concentrations, as confirmed by a 
strong correlation with independent MISR satellite measurements of aerosol concentrations at blue wavelengths ($\lambda=443$ nm).
However, this correlation might be limited by the low statistics arising from 
the small temporal overlap of H.E.S.S. and MISR observations as well as the
limited amount of H.E.S.S. data with very low values of $T$.
Furthermore, this quantity is not able to distinguish between large-scale clouds and 
aerosols and is based on some simplified assumptions,
such as the perfect inverse proportionality between telescope energy threshold and the muon efficiency. Addressing these points might result in a better correlation between
the atmospheric absorption measured by satellites and the
Cherenkov transparency coefficient. Simultaneous observations of on-site radiometer and LIDAR data and the Cherenkov telescope might also help and are currently under study \cite{bib:Vasileiadis} \cite{bib:Chadwick}.

The Cherenkov transparency coefficient is currently used
as a data quality parameter in H.E.S.S. Previous methods,
using other atmosphere-sensitive parameters \cite{bib:HEGRA} \cite{bib:VERITAS} \cite{bib:MAGIC} \cite{bib:Nolan}, have been used to correct the flux for changes
in atmospheric conditions. The strong correlation with independent atmospheric measurements suggests that the
Cherenkov transparency coefficient could be applied in the
same way, currently under study, making it possible to use
the Cherenkov technique over a wider range of atmospheric conditions. 

Furthermore, since all the parameters needed to derive $T$ are available from
routine calibration and quality checks that are generic to the IACT technique,
 in principle this quantity can be also implemented in other IACT experiments.
Its implementation in the future Cherenkov Telescope Array CTA is also under study.

\vspace*{0.5cm}
\footnotesize{{\bf Acknowledgment: }{The authors would like to acknowledge
the support of their host institutions. We want to thank the whole
H.E.S.S. collaboration for their support, especially Prof. Werner
Hofmann and Prof. Christian Stegmann as well as Prof. Thomas
Lohse and Dr. Ira Jung for the many fruitful discussions.}}

\end{document}